\documentclass[prl, twocolumn, showpacs,amsmath,amssymb]{revtex4}

\usepackage{wasysym}
\usepackage{graphicx}
\usepackage{dcolumn}
\usepackage{bm}
\usepackage{xcolor}
\newcommand{\texp}{T_\text{exp}}
\newcommand{\tof}{T_\text{tof}}

\begin{document}

\title{Backscattering echo of correlated wave packets}

\author{Nicolas Cherroret and Dominique Delande}
\affiliation{Laboratoire Kastler-Brossel, UPMC-Paris 6, ENS, CNRS; 4 Place Jussieu, F-75005 Paris, France}

\date{\today}

\begin{abstract}
We present an analytical theory describing a novel phenomenon of enhanced backscattering corresponding to the spatial refocussing of a spread-out correlated wave packet due to a brief interaction with a disordered potential. Our theory is validated by numerical simulations and explains the physics observed in recent experiments on cold atom transport in disorder.
\end{abstract}


\maketitle
 
\emph{Introduction}\indent When propagating in a disordered environment, a wave is scattered from the hilly landscape and undergoes successive elastic scattering processes \cite{Sheng95}. In certain circumstances, the probability for the wave to be backscattered can be strongly enhanced due to a great variety of phenomena. One well-known example is coherent backscattering, which is a constructive interference effect between two partial wave paths propagating in opposite directions in the regime of multiple scattering \cite{Akkermans86}. Coherent backscattering requires a spatially-coherent source and has been widely studied with various kinds of classical waves \cite{Aegerter09}, and recently with matter waves \cite{Jendrzejewski12, Labeyrie12}. In optics, other phenomena leading to enhanced backscattering have been identified as well. Among them one can think of the glory effect, which corresponds to a halo that is sometimes seen in a cloudy atmosphere around the shadow of an observer \cite{Bryant74}. As coherent backscattering, the glory is an interference effect, but it occurs in the single scattering regime and stems from the interference between wave paths inside droplets of water. Another effect responsible for enhanced backscattering is the opposition surge, well known in astrophysics. It is associated with a maximum of  brightness of a rough surface when the source of light is directly behind the observer \cite{Oetking66}, and has been in particular observed for the moon \cite{Gehrels56}. Unlike the two previous examples however, the opposition surge is a classical effect, i.e. it does not rely on the coherence of the source.

In a recent transport experiment using cold atoms \cite{Labeyrie12}, the enhanced backscattering of a matter wave packet (a Bose-Einstein condensate) was studied in a configuration where the packet first expands  with a mean momentum, then interacts shortly with a disordered optical potential and finally is detected after time of flight. In this experiment, three mechanisms yielding enhanced backscattering in the density distribution of the matter wave were identified. First an effect similar to the opposition surge, observable at short times for an average amplitude of the disordered potential comparable to the atomic kinetic energy. Second the coherent backscattering effect, and third a novel phenomenon, an echo presumably due to a refocussing of atoms backscattered by the disorder. This \emph{backscattering echo} (BSE) was visible only at finite time of flight and was conjectured to originate from the strong position-momentum correlations developed by the wave packet during its initial expansion. 

In this Letter we develop an analytical theory explaining the physics of the BSE. This theory is validated by numerical simulations and explains the experimental observations of \cite{Labeyrie12}. From a theoretical point of view, it provides a unified framework for describing the dynamics of a wave packet that propagates in a disordered potential and is then detected at \emph{arbitrary} time of flight. From a practical point of view, our work also suggests an original method to spatially refocus a wave packet by means of elastic reversal of particles momenta, which can be of high interest in the context of cold atom transport in random optical potentials \cite{Clement06, Shapiro0712} or more generally for the manipulation or shaping of matter wave packets.

\begin{figure}[h!]
\includegraphics[width=8.4cm]{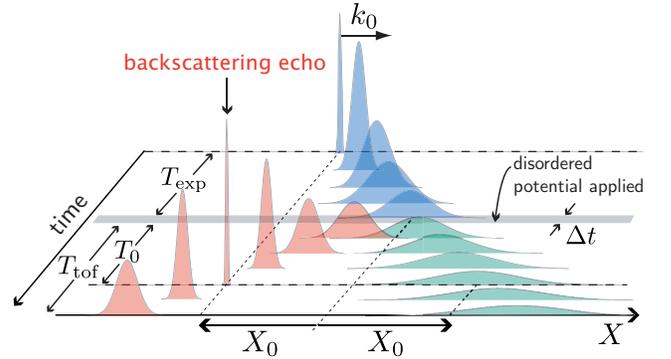}
\caption{\label{Scheme_BSE} 
(color online). Principle of BSE in 1D. A wave packet first expands freely from $t=0$ to $\texp$ (blue profiles), thereby developing position-momentum correlations. It is then subjected to a disordered potential for a short duration $\Delta t$. When the latter is turned off, the wave packet experiences a free flight of duration $T_\text{tof}$. At $T_\text{tof}=T_0\simeq\texp$, particles scattered forward (in green) generate a broad density profile, while backscattered particles (in red) refocus: this is the BSE effect.
}  
\end{figure}

\emph{Qualitative picture of BSE}\indent The principle of backscattering echo is illustrated in Fig. \ref{Scheme_BSE}, in 1D for simplicity. A wave packet is released at $t=0$ with a mean momentum $k_0$ and starts an expansion for a duration $\texp$, thereby developing strong position-momentum correlations (i.e. particles acquire a velocity proportional to their distance from the wave-packet center). We here forget interactions between particles. They do not play an essential role in the physics of the BSE and will be discussed later on.
Suppose now that at $t=\texp$ a disordered potential is applied for a short duration $\Delta t$. Particles are scattered either in the forward direction $+x$ or in the backward direction $-x$, and if at $\texp+\Delta t$ the disordered potential is turned off, backscattered particles (depicted in red in Fig. \ref{Scheme_BSE}) all refocus and reform the initial wave packet after a certain  time of flight $\tof=T_0$: this is the BSE effect. On the other hand, the part of the wave packet containing particles scattered in the forward direction keeps expanding freely and produces a broad density profile (depicted in green in Fig. \ref{Scheme_BSE}). As we will show below, this scenario remains valid in 2D, although the BSE peak now develops around a ring. The BSE time $T_0$ can be estimated by simply seeing the refocussing process as a time-reversed wave-packet expansion: $T_0\simeq \texp\simeq\Delta x/(\hbar\Delta k_x/m)$, where $\Delta x$ ($\Delta k_x$) is the spatial (momentum) dispersion of the wave packet at $t=\texp$. 

\emph{Theory}\indent We now calculate explicitly the density distribution of the wave packet after time of flight within the scenario described above, in a more general 2D configuration. It is important to note that since position-momentum correlations are crucial for producing the BSE effect, we must work in \emph{phase space} and thus resort to the formalism of the Wigner function. 
The calculation is done in three steps. First we determine the Wigner function $W(\textbf{r},\textbf{k},\texp)$ at the end of the initial stage of expansion. Then we establish the relation between the disorder-averaged Wigner function after propagation in the disorder, $\overline{W}(\textbf{r},\textbf{k},\texp+\Delta t)$, and $W(\textbf{r},\textbf{k},\texp)$. Finally from $\overline{W}$ we calculate the wave-packet density after a time of flight $\tof$.

At $t=0$ the wave packet has a mean momentum $\textbf{k}_0$ and its wave function has a certain envelope that we assume to be Gaussian, $\psi(\textbf{r},t=0)=1/\sqrt{\pi r_0^2}\exp\left[-\textbf{r}^2/(2 r_0^2)+i\textbf{k}_0.\textbf{r}\right]$. The packet thus starts moving and expanding freely, with a (boosted) wave function at time $t>0$ given by
\begin{eqnarray}
\label{phi_t}
\psi(\textbf{r},t)=\dfrac{e^{i\beta(t)}}{b(t)}
\psi\left[\dfrac{\textbf{r}-\hbar t\textbf{k}_0/m}{b(t)},0\right]
e^{\frac{i m \left(\textbf{r}-\frac{\hbar t\textbf{k}_0}{m}\right)^2}{2\hbar}\frac{\dot b(t)}{b(t)}}
\end{eqnarray}
where $\beta(t)$ is a real phase (irrelevant for our purpose) and $b(t)=\sqrt{1+\hbar^2 t^2/(m^2 r_0^4)}$. After an expansion time $t=\texp$,
the Wigner function $W(\textbf{r},\textbf{k},\texp)\equiv
\int d^2\!\boldsymbol{\rho}\, e^{-i \textbf{k}\cdot\boldsymbol{\rho}}
\psi(\textbf{r}+\boldsymbol{\rho}/2,\texp)\psi^*(\textbf{r}-\boldsymbol{\rho}/2,\texp)$ takes the simple form
\begin{eqnarray}
\label{W_ti}
W(\textbf{r},\textbf{k},\texp)=
4e^{
-r_0^2\left(\textbf{k}-\textbf{k}_0\right)^2
-\left(\textbf{r}-\frac{\hbar \texp\textbf{k}}{m}\right)^2/r_0^2
}.
\end{eqnarray}
Eq. (\ref{W_ti}) clearly emphasizes the  position-momentum correlations generated during the initial expansion, also evidenced from the inequality $\Delta r\Delta k> 1/2$, where $\Delta r=(r_0/\sqrt{2})\sqrt{1+(\hbar \texp/m r_0^2)^2}$ and $\Delta k=1/(\sqrt{2} r_0)$ are the radial spatial and momentum widths of the packet, respectively.

At $t=\texp$ the disordered potential is turned on for a duration $\Delta t$ (see Fig. \ref{Scheme_BSE}). Particles are scattered elastically in all directions, which leads to a spreading of the the wave packet around its mean position $\textbf{R}_0=\hbar \textbf{k}_0\texp/m$. Within a first, simple description, the disorder-averaged Wigner function after spreading is expressed as
\begin{eqnarray}
\label{W_tipdt}
&&\overline{W}(\textbf{r},\textbf{k},\texp+\Delta t)=\\\nonumber
&&\int d^2\textbf{r}'\int\! \frac{d^2\textbf{k}'}{(2\pi)^2}
\frac{\delta(\epsilon_k-\epsilon_{k'})}{\nu_0}\delta(\textbf{r}-\textbf{r}')W(\textbf{r}',\textbf{k}',\texp),
\end{eqnarray}
where $\nu_0=m/(2\pi\hbar^2)$ is the 2D density of states per unit surface and $\epsilon_k=\hbar^2\textbf{k}^2/(2m)$. Eq. (\ref{W_tipdt}) relies on three approximations: 1-- weak disorder $k_0\ell\gg 1$ with $\ell$ the mean free path associated with the disorder,  2-- diffusion approximation $\Delta t\gg\tau$ with $\tau$ the mean free time and 3-- $\sqrt{D\Delta t}\ll\Delta r$ with $D$ the diffusion coefficient \cite{AM}. Due to condition 1, interference phenomena such as Anderson localization are negligible \cite{SP07, Skipetrov08,Karpiuk12}, and in addition scattering from the disorder is rigorously energy conserving, as described by the first delta function in Eq. (\ref{W_tipdt}) \cite{Shapiro0712}. Condition 2 implies that the expansion process in the disorder produces a complete isotropization of particles momenta \cite{AM}. Finally, due to condition 3 the typical scale $|\textbf{r}-\textbf{r}'|\sim\sqrt{D\Delta t}$ over which particles move in the disorder is small as compared to the size $\Delta r$ of the wave packet, such that the transport propagator from $\textbf{r}'$ to $\textbf{r}$ in disorder can be approximated by $\delta(\textbf{r}-\textbf{r}')$.


In a last stage, at $t=\texp+\Delta t$ the disorder is turned off and particles experience a free flight of duration $\tof$, after which the disorder-averaged wave-packet density at time $\texp+\Delta t+\tof$ reads
\begin{eqnarray}
\label{final_Psi2}
n(\textbf{r},\tof)=\int\!
\frac{d^2\textbf{k}}{(2\pi)^2}
\overline{W}\left(\textbf{r}-\frac{\hbar \tof}{m}\textbf{k},\textbf{k},\texp+\Delta t\right)
\end{eqnarray}
[we write $n(\textbf{r},\tof)$ instead of $n(\textbf{r},\tof+\texp+\Delta t)$ for brevity]. Eq. (\ref{final_Psi2}) with $\overline{W}$ given by Eq. (\ref{W_tipdt}) is the general expression for the wave-packet density at any time of flight. Before addressing the interesting case $\tof=T_0=\Delta r/(\hbar\Delta k/m)$ where the BSE is expected, it is instructive as a first check to examine the limits $\tof=0$ and $\tof\rightarrow\infty$. First for $\tof=0$ we straightforwardly obtain $n(\textbf{r},0)\simeq |\psi(\textbf{r},\texp)|^2$ from Eq. (\ref{final_Psi2}). This simple result indicates that the wave-packet profile is practically not altered by the propagation in the disorder, which is a logical consequence of our approximation 3.
In the limit $\tof\rightarrow\infty$ on the other hand, we find
\begin{equation}
\label{infiniteT}
n(r,\infty)\sim \int \frac{d\theta}{2\pi}\left|\tilde\psi\left(k=\frac{mr}{\hbar \tof},\theta,\texp\right)\right|^2,
\end{equation}
where $\tilde\psi(k,\theta,t)$ denotes the Fourier transform of $\psi(\textbf{r},t)$ [given by Eq. (\ref{phi_t})] with $\theta$ is the angle between $\textbf{k}$ and $\textbf{k}_0$. Eq. (\ref{infiniteT}) parameterizes a ring of mean radius $\hbar |\textbf{k}_0| \tof/m$ and radial width $\hbar\Delta k \tof/m$. This ring mimics the  \emph{momentum} distribution of the wave packet, and stems from the isotropization of the momentum distribution $|\tilde\psi(\textbf{k}',\texp)|^2$ due to multiple scattering in the disorder, a phenomenon that has been recently demonstrated both theoretically \cite{Cherroret12} and experimentally \cite{Jendrzejewski12, Labeyrie12} with cold atoms.

\begin{figure}[h]
\includegraphics[width=7.3cm]{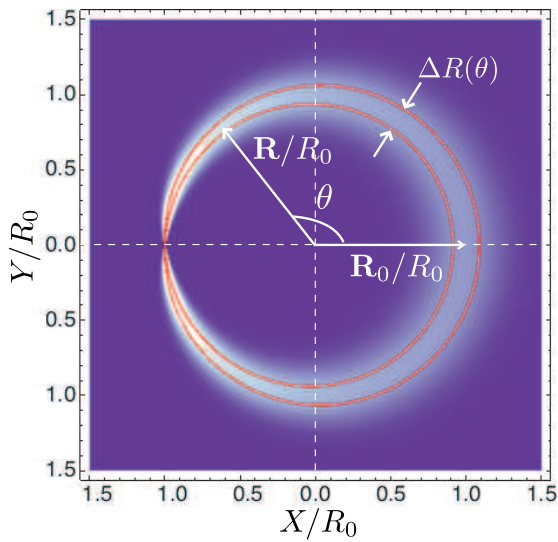}
\caption{\label{Ring_analysis} 
(color online). Density plot of Eq. (\ref{Psi2_BSE}): the 2D density distribution of the wave packet at $\tof=T_0$ has a ring shape, and is squeezed and enhanced around the backscattering direction $\theta=\pi$ (BSE effect). The two red curves delimit the mean ring width $\Delta R(\theta)=\Delta r\sqrt{2(1+\cos\theta)}$.}  
\end{figure}

We now evaluate Eq. (\ref{final_Psi2}) for $\tof=T_0$, i.e. at the BSE time (see Fig \ref{Scheme_BSE}). It is convenient to express the result in the coordinate system $\textbf{R}=\textbf{r}-\textbf{R}_0$ where the position is measured with respect to the center of mass of the wave packet, $\textbf{R}_0=\hbar \textbf{k}_0\texp/m$. Within the same approximations as above, we find
\begin{eqnarray}
\label{Psi2_BSE}
n\left(\textbf{R},T_0\right)\simeq
\dfrac{\exp\left[-\dfrac{(\textbf{R}^2-\textbf{R}_0^2)^2}{8\Delta r^2(\textbf{R}+\textbf{R}_0)^2}\right]}{(2\pi)^{3/2}\Delta r|\textbf{R}+\textbf{R}_0|}.
\end{eqnarray}
Eq. (\ref{Psi2_BSE}) is the 2D analog of the 1D (green+red) density profile sketched in Fig. \ref{Scheme_BSE} (with the correspondences $\textbf{R}\leftrightarrow X$ and $\textbf{R}_0\leftrightarrow X_0$), and fulfills the normalization condition $\int d^2\textbf{R}\,n\left(\textbf{R},T_0\right)=1$. We show in Fig. \ref{Ring_analysis} a density plot of Eq. (\ref{Psi2_BSE}) in the plane $(X/R_0, Y/R_0$), with $X=R\cos\theta$, $Y=R\sin\theta$ and $R_0\equiv |\textbf{R}_0|$.
The figure displays a clear enhancement of the distribution at $\theta=\pi$, i.e. a spatial refocussing at $\textbf{r}=0$ which is the 2D counterpart of the BSE effect shown in Fig. \ref{Scheme_BSE}. Note that a radial average of Eq. (\ref{Psi2_BSE}) leads to a uniform angular profile and thus to a suppression of the BSE effect, which validates the method used in \cite{Labeyrie12} to observe coherent backscattering. The squeezing seen around $\theta=\pi$ is evidenced from the calculation of the mean ring width $\Delta R(\theta)=\Delta r\sqrt{2(1+\cos\theta)}$. The boundaries $R_0\pm\Delta R(\theta)/2$ are shown in Fig. \ref{Ring_analysis} as red curves. 

At this stage two important comments are in order. First, the result $\Delta R(0)=2\Delta r$ indicates a doubling of the size of the packet between $t=\texp$ and $t=\texp+T_0\simeq 2\texp$, as expected for a ballistic expansion. Second, the value $\Delta R(\pi)=0$ signals a perfect refocussing, associated with a divergence of the distribution (\ref{Psi2_BSE}). This idealized behavior underlines the breakdown of Eq. (\ref{Psi2_BSE}) around $\theta=\pi$, which stems from the neglect of all possible sources of spatial broadening of the BSE peak in our calculation, symbolized by the two delta functions in Eq. (\ref{W_tipdt}). In what follows we improve on this approximation.

In a real experimental configuration $\Delta R(\pi)$ cannot be exactly zero because a number of effects may alter position-momentum correlations of the wave packet and broaden the BSE peak. An important source of broadening is due to the finite propagation time $\Delta t$ in the disorder. It can be accounted for by replacing the second delta function in Eq. (\ref{W_tipdt}) by the diffusion propagator $\exp[-|\textbf{r}-\textbf{r}'|^2/(4D\Delta t)]/(4\pi D \Delta t)$ \cite{AM}. In this case we find that at $\theta =\pi$, Eq. (\ref{Psi2_BSE}) should be corrected to
\begin{eqnarray}
\label{BSE_broadening}
n(R,\theta=\pi,T_0)\simeq \dfrac{\exp\left[-\dfrac{(R-R_0)^2}{4 D \Delta t}\right]}{(2\pi)^{3/2} R_0\sqrt{2D\Delta t}},
\end{eqnarray}
with the definition $n(R,\theta,T_0)\equiv n(\textbf{R},T_0)$. This shows that the finite propagation time in the disorder induces a broadening $\sim\sqrt{D\Delta t}$. Comparing Eq. (\ref{BSE_broadening}) with Eq. (\ref{Psi2_BSE}), we define the contrast of the BSE peak as the ratio $n(R_0,\pi,T_0)/n(R_0,0,T_0)=\sqrt{2}\Delta r/\sqrt{D\Delta t}$. We also estimate the angular width of the peak at radius $R_0$, $\Delta\theta_\text{BSE}\simeq4\sqrt{2D\Delta t}/\Delta r$, defined as twice the angle at which Eq. (\ref{Psi2_BSE}) equals half of the maximum of Eq. (\ref{BSE_broadening}).

\emph{Numerical simulations}\indent To test the validity of our theory, we perform 2D classical numerical simulations of particle transport, starting from the Gaussian phase-space distribution ($\ref{W_ti}$) and using a speckle random potential as in recent experiments on cold atoms \cite{Clement06}. The length unit is the correlation length $\zeta$ of the random potential, which defines the characteristic time $\tau_\zeta=\hbar^2/(m\zeta^2)$ and energy $E_\zeta=\hbar^2/(m\zeta^2)$ scales \cite{Kuhn07}. In these units, we choose a random potential amplitude $V=0.1$, an initial momentum $\textbf{k}_0=(1,0)$ (in cartesian coordinates), a spatial $\sqrt{2}\Delta r=400$ and momentum $\sqrt{2}\Delta k=0.1$ dispersion (note that $\Delta r\Delta k\gg1/2$) and a propagation time in disorder $\Delta t=900$. 
\begin{figure}[h!]
\includegraphics[width=8.2cm]{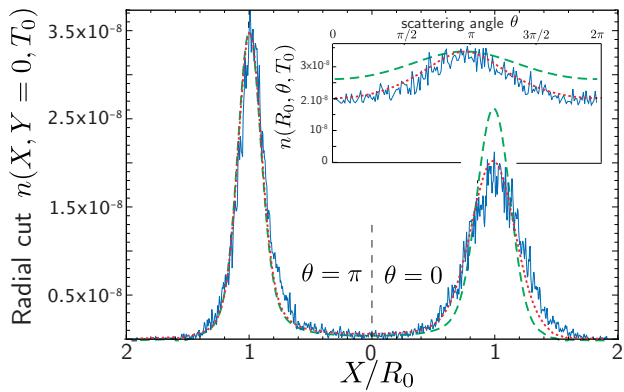}
\caption{\label{BSE_profile} 
(color online). Main plot: radial cut of $n(\textbf{R},T_0)$ along $Y=0$. Inset: angular profile $n(R_0,\theta,T_0)$. Solid curves (blue online) are numerical simulations for a Gaussian initial wave function. Dotted curves (red online) are plots of Eq. (\ref{final_Psi2}). Dashed curves (green online) show the theoretical prediction for a Thomas-Fermi initial wave function (it practically coincides with the result for a Gaussian wave function at $\theta=\pi$).
}  
\end{figure}
We show in the main plot of Fig. \ref{BSE_profile} a numerical radial cut along $Y=0$ of $n(\textbf{R},T_0)$ 
as a function of $X/R_0$, and in the inset a numerical angular cut at $R=R_0$, i.e. $n(R_0,\theta,T_0)$, both obtained after averaging over 960 disorder realizations. The BSE effect is well visible. For comparison we also plot the profile obtained from Eq. (\ref{final_Psi2}) (dotted curve, red online). In evaluating (\ref{final_Psi2}) we use the exact diffusion propagator and also account for the energy broadening due to disorder, i.e. we replace $\delta(\epsilon_k-\epsilon_{k'})$ by the associated spectral function in the weak disorder limit, $2\hbar\gamma/[(\epsilon_k-\epsilon_{k'})+(\hbar\gamma)^2]$ \cite{AM}, with $\gamma\simeq22$ a fit parameter whose value is consistent with the prediction 25.7 of weak-disorder perturbation theory \cite{Kuhn07}. We see that the agreement with the simulations is excellent.

The scenario presented in this Letter is well adapted to an experiment of atomic matter-wave transport in a random optical potential. In that concrete situation, two important ingredients might potentially affect the BSE. The first one is the shape of the wave-packet density at $t=0$, which is generally closer to an inverted parabola (Thomas-Fermi profile) than to a Gaussian for an atomic wave packet initially released from a harmonic trap \cite{Clement06}. In Fig. \ref{BSE_profile} we show as dashed curves (green online) the prediction of our theory for a Thomas-Fermi wave function. We see that while the global shape of the squeezed ring is slightly modified with respect to the Gaussian case, the BSE peak is not affected at all. This will remain true in general as long as the width of the initial wave function is smaller than the broadening $\sqrt{D\Delta t}$ induced by the disorder. The second one is the possible presence of repulsive atomic interactions. For an initial matter wave trapped in a harmonic potential of frequency $\omega$, all the approach presented here remains valid provided the replacement $b(t)\leftarrow\sqrt{1+(\omega t)^2}$ is performed in Eq. (\ref{phi_t}) \cite{Kagan96, Castin96}. For a given expansion time $\texp$, the essential effect of interactions is then to increase $\Delta r$ and $\Delta k$ with respect to the free case, thus altering the BSE time $T_0=\Delta r/(\hbar\Delta k/m)$.

To finish, let us compare the BSE with coherent backscattering (CBS). The BSE is a purely classical phenomenon whereas CBS stems from the quantum interference of counter-propagating wave-path amplitudes \cite{Akkermans86, Aegerter09}. In an experiment involving both a coherent \emph{and} correlated wave packet and where the density is recorded after time of flight, both BSE and CBS come together \cite{Labeyrie12}. They have nevertheless different signatures: 1-- the angular width $\Delta \theta_\text{CBS}$ of the CBS peak is narrow, whereas the angular width $\Delta \theta_\text{BSE}$ of the BSE peak is rather broad, 2-- BSE is accompanied with a squeezing of the density profile, 3-- $\Delta \theta_\text{CBS}\sim\lambda\sqrt{D\Delta t}^{-1}$ \cite{Cherroret12} whereas $\Delta \theta_\text{BSE}\sim\sqrt{D\Delta t}/\Delta r$ and 4-- CBS is sensitive to processes breaking time-reversal invariance unlike BSE.

\emph{Conclusion}\indent We have theoretically described a novel phenomenon of enhanced backscattering due to disorder, a backscattering echo which can be observed in experiments involving wave packets with position-momentum correlations. Our theory sheds new light on recent experiments on cold atoms \cite{Labeyrie12, Jendrzejewski12} and provides further insight for future studies of wave-packet transport in disorder.

We thank Christian Miniatura and Guillaume Labeyrie for interesting discussions, and acknowledge financial support from the French ANR (Project No. 11-B504-0003 LAKRIDI). Part of this work was performed using HPC resources from GENCI- [CCRT/CINES/IDRIS]  (Grant 2012- [i2012056089]).

\end{document}